\documentclass[showpacs,preprintnumbers,amsmath,amssymb,twocolumn]{revtex4}

\usepackage{amssymb}

\usepackage{amsmath}
\usepackage{dcolumn}
\usepackage{bm}
\usepackage{graphicx}
\usepackage{color}

\begin{document}

\title{Microcanonical phase transitions in small systems}
\author{Michele Campisi}
\email{campisi@unt.edu} \affiliation{Department of
Physics,University of North Texas Denton, TX 76203-1427, U.S.A.}
\date{\today }

\begin{abstract}
When studying the thermodynamic properties of mesoscopic systems
the most appropriate microcanonical entropy is the volume entropy,
i.e. the logarithm of the volume of phase space enclosed by the
hypersurface of constant energy. For systems with broken
ergodicity, the volume entropy has discontinuous jumps at values
of energy that correspond to separatrix trajectories.
Simultaneously there is a convex intruder in the entropy function
and a region of negative specific heat below such critical
energies. We illustrate this with a simple model composed of a
chain of 3 particles which interact via a Lennard-Jones potential.
\end{abstract}

 \maketitle

\section{\label{sec:intro}Introduction}

The work of D.H.E. Gross \cite{Gross97} has recently pointed out
the attention on the fact that a microcanonical description of
systems which may display phase transitions is in general more
adequate than the traditional canonical one. This is because the
canonical description may ``smear out'' important information
contained in the microcanonical description which is richer
\cite{Gross97}. For example negative specific heats, which have
recently been observed experimentally in mesoscopic systems
\cite{PhysRevLett.86.1191,PhysRevLett.87.203401,Gobet02}, can be
accounted for in the microcanonical ensemble but not in the
canonical one \cite{Lynden-bell99}. Indeed it is well known that
canonical ensemble and microcanonical ensemble are not in general
equivalent, even when the thermodynamic limit is considered
\cite{Gross97,Gallavotti,Ruffo05,Touchette04}.

The statistical mechanical analysis of physical systems based on
canonical ensemble is quite well established and universally
agreed upon. Roughly speaking one has to compute the partition
function $Z(\beta)$ and derive the thermodynamics of the system
from the free energy $F = -\beta^{-1}\ln Z(\beta)$. Things are not
quite so broadly agreed upon in the case of the microcanonical
ensemble. In fact since the pioneering works of Boltzmann and
Gibbs two possibilities were given for the microcanonical analysis
of physical systems, which correspond to two different definitions
of entropy (see for example the textbook of Gibbs \cite{Gibbs02}
or the more recent textbook of Huang \cite{Huang}). Following
\cite{Campisi05} and \cite{adib04} we shall refer to these two
entropies as ``surface entropy'' and ``volume entropy''. The
surface entropy is defined as:\footnote{In this work we adopt the
convention that Boltzmann constant $k_B$ is equal to $1$.}
\begin{equation}\label{eq:SurfaceEntropy}
    S_\Omega(E) = \ln \Omega(E)
\end{equation}
where \begin{equation}\label{}
    \Omega(E) = \int d\mathbf{z} \delta(E-H(\mathbf{z}))
\end{equation}
with $\delta(x)$ denoting Dirac delta function. Sometimes this is
also referred to as Boltzmann entropy or Boltzmann-Planck entropy.
The volume entropy is defined as:
\begin{equation}\label{}
    S_\Phi(E) = \ln \Phi(E)
\end{equation}
where
\begin{equation}\label{}
    \Phi(E) = \int d\mathbf{z} \theta(E-H(\mathbf{z}))
\end{equation}
with $\theta(x)= \int_{-\infty}^{x}dy\delta(y)$ denoting Heaviside
step function. The quantities $\Omega$ and $\Phi$ are related
through the differential equation \cite{Khinchin}:
\begin{equation}\label{}
    \Phi'(E) = \Omega(E)
\end{equation}
where the prime symbol denotes derivation with respect to $E$.
These entropies are named surface entropy and volume entropy
because they are calculated as the logarithm of the area of the
hyper-surface of constant energy in phase space and the volume of
phase space that it encloses respectively.

With reference to the literature about microcanonical phase
transitions the surface entropy is certainly the most popular. For
example Barr\'{e} \emph{et. al.} \cite{Ruffo05} method based on
large deviation techniques uses the surface entropy. The surface
entropy is used also in Rugh's microcanonical formalism adopted in
Ref. \cite{Campa06}. The strongest advocate of surface entropy is
perhaps Gross \cite{gross:224111}. Nonetheless pioneers of
microcanonical phase transitions, such as Thirring
\cite{Thirring70} and Lynden-Bell \cite{Lynden-bell99}, used the
volume entropy.

The two entropies coincide in the thermodynamic limit but when the
number of degrees of freedom of the system under study is small
relevant differences may appear, therefore it is necessary to
choose properly. Some authors \cite{Naudts05,adib04} have already
pointed out that the surface entropy is not adequate when dealing
with small systems because it does not account properly for
finite-size effects. On the other hand there is a number of
theoretical reasons to prefer the volume entropy when the number
of degrees of freedom becomes small. Here we shall review these
reasons and we will illustrate the employment of the volume
entropy in the study of phase transition with a small
Lennard-Jones chain which displays a region of negative heat
capacity.

\section{\label{sec:justification}Why volume entropy}
In this section we will summarize some old and recent results
concerning the volume entropy. These results indicate that, no
matter the number of degrees of freedom of the system under study,
the volume entropy always provides a good mechanical analogue of
thermodynamic entropy. For example Helmholtz \cite{Helmholtz}
proved that the logarithm of the area enclosed in phase space by
the trajectory of a 1-dimensional system (i.e., the 1D volume
entropy $S_\Phi$) provides a mechanical analogue of physical
entropy in the sense that if one considers the quantities $P
\doteq \langle\frac{\partial H}{\partial V}\rangle_t$ and $T
\doteq \langle 2 K \rangle_t$ where $K$ is the kinetic energy, $V$
is an external parameter on which the Hamiltonian depends and
$\langle \cdot \rangle_t$ denotes time average, then,
\begin{equation}\label{eq:heatTheo}
    \frac{dE+PdV}{T} = \text{exact differential} = dS_\Phi
\end{equation}
This result, known as Helmholtz Theorem
\cite{Gallavotti,Campisi05}, says that the volume entropy is a
good mechanical analogue of thermodynamic entropy in the sense
that it reproduces exactly the fundamental law of thermodynamics
(i.e. the \emph{heat theorem} (\ref{eq:heatTheo})). The Helmholtz
Theorem has been recently generalized to multi-dimensional ergodic
(i.e. metrically indecomposable) systems (see Ref.
\cite{Campisi05} or Ref. \cite{Cardin04} for a different but
equivalent approach). The resulting Generalized Helmholtz Theorem
essentially states that the volume entropy reproduces the heat
theorem \emph{no matter the number of degrees of freedom $N$}. The
same cannot be said about surface entropy which has been proved to
reproduce it only up to corrections of the order $O(1/N)$
\cite{Gallavotti}. As a matter of fact Gibbs presents the volume
entropy  in his celebrated \emph{Principles of Statistical
Mechanics} \cite{Gibbs02} as the entropy that naturally satisfies
the fundamental principle of thermodynamics (that is the heat
theorem) \cite{Uffink06}.

Hertz \cite{Hertz} pointed out that the volume entropy is an
adiabatic invariant already in 1910, and based his approach to
statistical mechanics on it. Among the textbooks that adopt the
same approach, those of M\"{u}nster \cite{Munster}, Becker
\cite{Becker69}, and the more recent book of Berdichevsky
\cite{Berdichevsky} are worth mentioning. Adiabatic invariance is
another good property of volume entropy because it reproduces
quite well Clausius' requirement that ``For every quasi static
process in a thermally isolated system which begins and ends in an
equilibrium state, the entropy of the final state is equal to that
of the initial state'' \cite{Uff01}. Of course the surface entropy
is not an adiabatic invariant, although it becomes approximately
such as the number of degrees of freedom increases
\cite{Campisi05}.

Very recently it has been also proved that non-adiabatic
transformations occurring in isolated systems which are initially
in a state of thermal equilibrium always result in an increase of
the expectation value of the volume entropy
\cite{Campisi07.4,Tasaki00}. This result too does hold \emph{no
matter the number of degrees of freedom $N$} and cannot be proved
in general for surface entropy. Thus the volume entropy explains
quite satisfactorily also Clausius' law of entropy increase ``For
every non quasi static process in a thermally isolated system
which begins and ends in an equilibrium state, the entropy of the
final state is greater than that of the initial state''
\cite{Uff01}.

Recently more and more authors are becoming aware of the
theoretical value of volume entropy. For example, on the basis of
a Laplace transform technique for the microcanonical ensemble,
Pearson \emph{et. al.} \cite{Pearson85} reached the conclusion
that the volume entropy ``is the most correct definition for the
entropy, even though it is unimportant for any explicit numerical
calculation'', meaning that in the thermodynamic limit the
difference with surface entropy becomes negligible. On the other
hand for small system, such intrinsic correctness of the volume
entropy becomes very important. Adib \cite{adib04} argues that the
finite size corrections to surface entropy found in Ref.
\cite{BMWG} would be unneeded if the volume entropy were used
instead.

It is worth mentioning that the volume entropy has another
property that is particularly important for small systems which
have negative heat capacity, namely it is a naturally
\emph{nonextensive} entropy. According to Lynden-Bell
\cite{Lynden-bell99}, in fact, systems with negative heat capacity
are necessarily \emph{nonextensive}. The property of
nonextensivity of volume entropy follows directly from the
composition rule of enclosed volumes, $\Phi_i$ $i=1,2$ , of two
systems with total energy $E= E_1+E_2$ \cite{Khinchin}:
\begin{equation}\label{eq:convolution}
    \Phi(E) = \int dE_1 \Phi_1'(E_1)\Phi_2(E-E_1)
\end{equation}
which is not a simple multiplication but a form of convolution
which accounts for all possible partitions of energies between the
two systems. It has to be stressed that, despite of what is often
stated in literature, the surface entropy is nonextensive too, as
the composition rule for surface integrals $\Omega_i$ is the
convolution, not the multiplication \cite{Khinchin}:
\begin{equation}\label{eq:convolution}
    \Omega(E) = \int dE_1 \Omega_1(E_1)\Omega_2(E-E_1)
\end{equation}

 In sum, the volume entropy accounts for certain basic
principles of thermodynamics, like the heat theorem and Clausius
formulation of the second law equally well for large and small
systems, whereas the surface entropy accounts for them only in the
case of large system. For this reason it is the most appropriate
\emph{mechanical analogue of thermodynamic entropy} when dealing
with small systems.

\section{\label{sec:model}Lennard-Jones chain}
According to the Helmholtz Theorem \cite{Campisi05} the mechanical
analogue of thermodynamic entropy of a one dimensional system is
\begin{equation}
S_\Phi(E,V)=\log 2\int_{x_{-}\left( E,V\right) }^{x_{+}\left( E,V\right) }%
dx\sqrt{2m\left( E-\varphi \left( x,V\right) \right) }
\label{helmEnt}
\end{equation}
where $x_{\pm}\left( E,V\right)$ denote the turning points of the
trajectory. If the potential is such that there is only one
trajectory per energy level (ergodicity), then $S_\Phi$ satisfies
Eq. (\ref{eq:heatTheo}) \cite{Gallavotti,Campisi05}. Nonetheless,
if the system has more than one trajectory per energy level for a
certain range of energies and the system is on \emph{one} of them,
still we can use the above formula and say that the heat theorem
is satisfied as long as the energies considered lye within that
energy range. In this case $P$ and $T$ would be calculated as time
averages over the \emph{actual} trajectory and $S_\Phi$ would be
given by the area enclosed by \emph{that} trajectory only. Let us
illustrate this with a practical example. Let us consider a 1D
chain composed of three particles which interact via a
Lennard-Jones potential. Let us fix the position of two of them
and let us place the third one in between, so that the first two
particles act as walls of a 1D box. Let us now study the behavior
of the particle inside the box. Let the interaction potential be:
\begin{equation}\label{}
    u(x) = \frac{1}{x^{12}}-\frac{1}{x^{6}}
\end{equation}
and let us place the ``walls'' at $x=\pm V/2$. Then the particle
in the box is subject to the following potential:
\begin{equation}\label{}
    \varphi(x,V) = u(x+V/2)+u(x-V/2)
\end{equation}
For values of $V$ larger than a certain critical value $V_c \simeq
2.5$, this system has a critical energy $E_c(V) = \phi(0,V)$ such
that for energy below $E_c(V)$ ergodicity is broken and there are
two trajectories per energy level. Above $E_c$ the dynamics is
ergodic and there is only one trajectory per energy level.
\begin{figure}
\includegraphics[width=8 cm]{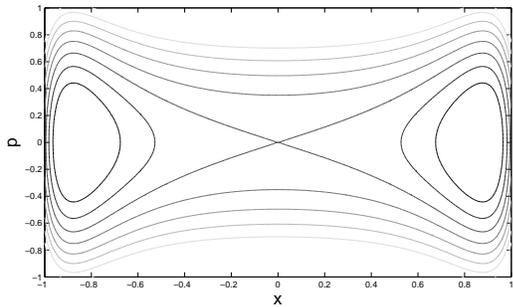}
  \caption{Phase space structure for a particle of mass $m=1$ in a Lennard-Jones box of size $V=4>V_c$. The separatrix corresponds to the critical energy
  $E_c=-0.0308$. Below the critical energy there are two distinct
  trajectories (the dynamics is not ergodic).}
\label{fig:phaseSpace}
\end{figure}
Figure \ref{fig:phaseSpace} shows a contour plot of various energy
levels in phase space for a particle of mass $m=1$ in the
Lennard-Jones box of size $V=4>V_c$ . For energy $E=E_c$ we have a
separatrix. Below $E_c$ the curve of constant energy splits into
two disconnected curves, whereas for values of $E$ larger than
$E_c$ we have only one curve. Below the critical energy the volume
integral $\Phi$ is given by the area enclosed by \emph{one} of the
two possible trajectories. As the energy crosses the critical
value the integral $\Phi$ jumps discontinuously. In formulae we
have:
\begin{eqnarray}\label{}
\Phi(E)= \left[\frac{1}{2} \theta(E_c-E) + \theta(E-E_c)\right] \nonumber\\
\times
 \int \left[2m(E-\varphi(x,V))\right]^{1/2}_+ dx
\end{eqnarray}
The symbol $[y]^{1/2}_+$ denotes a function that is equal to
$\sqrt{y}$ for $y \geq 0$ and is null otherwise. The volume
entropy, which is calculated by taking the logarithm of the
expression above, then displays a jump at the critical energy as
well.
\begin{figure}
\includegraphics[width=8 cm]{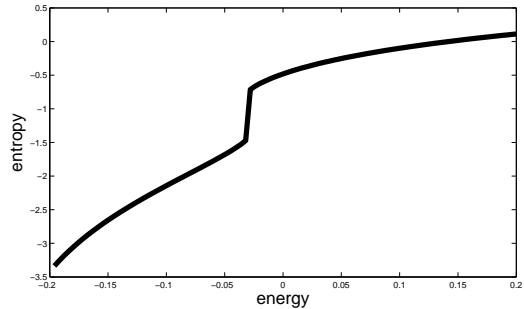}
  \caption{Entropy versus Energy for a particle of mass $m=1$ in a Lennard-Jones box of size $V=4>V_c$.
  The discontinuity of Entropy at the critical energy signals a discontinuous phase transition.}
\label{fig:entropy}
\end{figure}
Figure \ref{fig:entropy} shows a plot of $S$ as a function of $E$,
for the values $m=1$ and $V=4>V_c$. The critical energy is
$E_c=-0.0308$.
\begin{figure}
\includegraphics[width=8 cm]{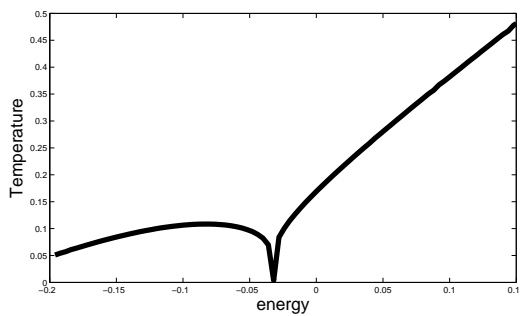}
  \caption{Temperature versus Energy for a particle in of mass $m=1$ in a Lennard-Jones box of size $V=4>V_c$.
  The curve displays a region of negative specific heat. At the critical energy the temperature goes to zero.}
\label{fig:temperature}
\end{figure}
Figure \ref{fig:temperature} shows the temperature plotted against
the energy calculated, according to Eq. (\ref{eq:heatTheo}), as
\footnote{This equation is essentially the microcanonical
equipartition theorem which is an exact result in classical
Hamiltonian mechanics \cite{Khinchin}}:
\begin{equation}\label{}
    T \doteq \left\langle 2K \right\rangle_t =
    \frac{\Phi(E)}{\Omega(E)}=\left(\frac{\partial S_\Phi}{\partial
    E}\right)^{-1}
\end{equation}
There is a region of negative slope in the graph which correspond
to a \emph{negative heat capacity}.

\section{\label{sec:discussion}Discussion}
\begin{figure}
\includegraphics[width=8 cm]{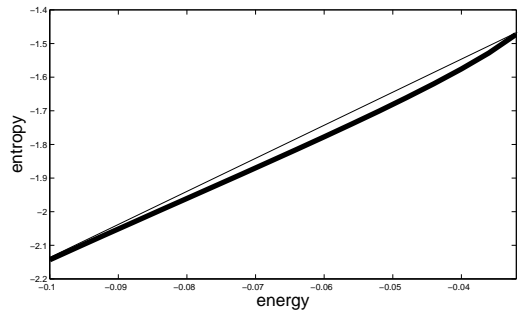}
  \caption{Entropy versus Energy for $E<E_c$. Right below the critical energy
  the entropy function (thick line) is convex (the thin straight line is only a guide for the eye).}
\label{fig:entropy-inset}
\end{figure}

The example provided in the previous section is perhaps too simple
to be of interest to any specific physical problem. Nonetheless it
illustrates qualitatively the mechanism of microcanonical phase
transition as captured by the volume entropy. Such phase
transitions are associated with the crossing of separatrix
trajectories, for which the dynamics of the system has no finite
time scale. The figures show neatly that at the separatrix energy
the entropy has a discontinuous jump, the temperature goes to
zero, and for energies below the critical energy we have a region
of negative heat capacity. These are not specific features of the
system studied\footnote{A similar behavior has been observed in a
chain of particles interacting via a $\phi^4$ potential studied
with volume entropy \cite{Cardin04} and has been predicted for the
pendulum \cite{Naudts05}.}. Whenever a separatrix is crossed there
is a sudden \emph{open-up} of a larger portion of phase space for
the trajectory to \emph{enclose} which leads to a discontinuity in
the entropy\footnote{The idea that microcanonical phase
transitions are associated to sudden open-up of phase space has
been expressed also in Ref. \cite{gross:224111}}. Further, at the
separatrix, the period of motion, which for a well known theorem
of classical mechanics is given by $\Phi'=\Omega$ \cite{LandauI},
becomes infinite. Therefore the temperature, \emph{i.e.}, $T=
\Phi/ \Omega$ goes to zero. Since the temperature $T \doteq
2<K>_t= <p^2/m>_t$ is definite positive, below the critical energy
there \emph{necessarily} is a region of negative slope, that is
negative heat capacity. The appearance of a negative heat capacity
is associated with a convex intruder in the entropy (see Fig.
\ref{fig:entropy-inset}) which signals the approach to the
separatrix from below.
It is important to notice that using the surface entropy would
lead to a \emph{drastically different} result. In this case the
temperature would be calculated as $T_\Omega = \Omega / \Omega'$,
which might not tend to zero at the critical energy! Note also
that $T_\Omega$ is not proportional to the average kinetic energy
and can be negative. Therefore, in agreement with Ref.
\cite{Naudts05} we believe that surface entropy is not suited for
low dimensional systems with broken ergodicity.

The volume entropy could be used to address microcanonical phase
transitions in small dimensional systems with either long or short
range interactions, like the $\phi^4$ model, chains of particles
interacting via Lennard-Jones potential \cite{Galgani72} or the
Hamiltonian Mean Field model \cite{Ruffo95}. All these models are
expected to undergo a breaking of ergodicity \cite{Galgani72,
Mukamel05}, thus there are separatrix trajectories and possible
phase transitions that the volume entropy can detect.

The advantage of using the volume entropy is that it provides a
good mechanical analogue of thermodynamic entropy even for small
system, thus accounting properly for the finite-size effects. As
the development of technology is allowing experimentalists to
probe the thermodynamic behavior of smaller and smallers systems,
this is becoming an increasingly important task. The main
limitation of the present approach is that it is restricted to
classical statistical mechanics, thus it does not account for
quantum-mechanical phenomena.


\begin{thebibliography}{33}
\expandafter\ifx\csname
natexlab\endcsname\relax\def\natexlab#1{#1}\fi
\expandafter\ifx\csname bibnamefont\endcsname\relax
  \def\bibnamefont#1{#1}\fi
\expandafter\ifx\csname bibfnamefont\endcsname\relax
  \def\bibfnamefont#1{#1}\fi
\expandafter\ifx\csname citenamefont\endcsname\relax
  \def\citenamefont#1{#1}\fi
\expandafter\ifx\csname url\endcsname\relax
  \def\url#1{\texttt{#1}}\fi
\expandafter\ifx\csname
urlprefix\endcsname\relax\def\urlprefix{URL }\fi
\providecommand{\bibinfo}[2]{#2}
\providecommand{\eprint}[2][]{\url{#2}}

\bibitem[{\citenamefont{Gross}(1997)}]{Gross97}
\bibinfo{author}{\bibfnamefont{D.~H.~E.} \bibnamefont{Gross}},
  \bibinfo{journal}{Physics Reports} \textbf{\bibinfo{volume}{279}},
  \bibinfo{pages}{119} (\bibinfo{year}{1997}).

\bibitem[{\citenamefont{Schmidt et~al.}(2001)\citenamefont{Schmidt, Kusche,
  Hippler, Donges, Kronm\"uller, von Issendorff, and
  Haberland}}]{PhysRevLett.86.1191}
\bibinfo{author}{\bibfnamefont{M.}~\bibnamefont{Schmidt}},
  \bibinfo{author}{\bibfnamefont{R.}~\bibnamefont{Kusche}},
  \bibinfo{author}{\bibfnamefont{T.}~\bibnamefont{Hippler}},
  \bibinfo{author}{\bibfnamefont{J.}~\bibnamefont{Donges}},
  \bibinfo{author}{\bibfnamefont{W.}~\bibnamefont{Kronm\"uller}},
  \bibinfo{author}{\bibfnamefont{B.}~\bibnamefont{von Issendorff}},
  \bibnamefont{and}
  \bibinfo{author}{\bibfnamefont{H.}~\bibnamefont{Haberland}},
  \bibinfo{journal}{Phys. Rev. Lett.} \textbf{\bibinfo{volume}{86}},
  \bibinfo{pages}{1191} (\bibinfo{year}{2001}).

\bibitem[{\citenamefont{Gobet et~al.}(2001)\citenamefont{Gobet, Farizon,
  Farizon, Gaillard, Buchet, Carr\'e, and M\"ark}}]{PhysRevLett.87.203401}
\bibinfo{author}{\bibfnamefont{F.}~\bibnamefont{Gobet}},
  \bibinfo{author}{\bibfnamefont{B.}~\bibnamefont{Farizon}},
  \bibinfo{author}{\bibfnamefont{M.}~\bibnamefont{Farizon}},
  \bibinfo{author}{\bibfnamefont{M.~J.} \bibnamefont{Gaillard}},
  \bibinfo{author}{\bibfnamefont{J.~P.} \bibnamefont{Buchet}},
  \bibinfo{author}{\bibfnamefont{M.}~\bibnamefont{Carr\'e}}, \bibnamefont{and}
  \bibinfo{author}{\bibfnamefont{T.~D.} \bibnamefont{M\"ark}},
  \bibinfo{journal}{Phys. Rev. Lett.} \textbf{\bibinfo{volume}{87}},
  \bibinfo{pages}{203401} (\bibinfo{year}{2001}).

\bibitem[{\citenamefont{Gobet et~al.}(2002)\citenamefont{Gobet, Farizon,
  Farizon, Gaillard, Buchet, Carr\'e, Scheier, and M\"ark}}]{Gobet02}
\bibinfo{author}{\bibfnamefont{F.}~\bibnamefont{Gobet}},
  \bibinfo{author}{\bibfnamefont{B.}~\bibnamefont{Farizon}},
  \bibinfo{author}{\bibfnamefont{M.}~\bibnamefont{Farizon}},
  \bibinfo{author}{\bibfnamefont{M.~J.} \bibnamefont{Gaillard}},
  \bibinfo{author}{\bibfnamefont{J.~P.} \bibnamefont{Buchet}},
  \bibinfo{author}{\bibfnamefont{M.}~\bibnamefont{Carr\'e}},
  \bibinfo{author}{\bibfnamefont{P.}~\bibnamefont{Scheier}}, \bibnamefont{and}
  \bibinfo{author}{\bibfnamefont{T.~D.} \bibnamefont{M\"ark}},
  \bibinfo{journal}{Phys. Rev. Lett.} \textbf{\bibinfo{volume}{89}},
  \bibinfo{pages}{183403} (\bibinfo{year}{2002}).

\bibitem[{\citenamefont{Lynden-Bell}(1999)}]{Lynden-bell99}
\bibinfo{author}{\bibfnamefont{D.}~\bibnamefont{Lynden-Bell}},
  \bibinfo{journal}{Physica A} \textbf{\bibinfo{volume}{263}},
  \bibinfo{pages}{293} (\bibinfo{year}{1999}).

\bibitem[{\citenamefont{Gallavotti}(1995)}]{Gallavotti}
\bibinfo{author}{\bibfnamefont{G.}~\bibnamefont{Gallavotti}},
  \emph{\bibinfo{title}{Statistical mechanics. A short treatise}}
  (\bibinfo{publisher}{Springer Verlag}, \bibinfo{address}{Berlin},
  \bibinfo{year}{1995}).

\bibitem[{\citenamefont{Barre}(May 2005)}]{Ruffo05}
\bibinfo{author}{\bibfnamefont{J.}~\bibnamefont{Barre}},
  \bibinfo{journal}{Journal of Statistical Physics}
  \textbf{\bibinfo{volume}{119}}, \bibinfo{pages}{677} (\bibinfo{year}{May
  2005}).

\bibitem[{\citenamefont{Touchette et~al.}(2004)\citenamefont{Touchette, Ellis,
  and Turkington}}]{Touchette04}
\bibinfo{author}{\bibfnamefont{H.}~\bibnamefont{Touchette}},
  \bibinfo{author}{\bibfnamefont{R.~S.} \bibnamefont{Ellis}}, \bibnamefont{and}
  \bibinfo{author}{\bibfnamefont{B.}~\bibnamefont{Turkington}},
  \bibinfo{journal}{Physica A} \textbf{\bibinfo{volume}{340}},
  \bibinfo{pages}{138} (\bibinfo{year}{2004}).

\bibitem[{\citenamefont{Gibbs}(1902)}]{Gibbs02}
\bibinfo{author}{\bibfnamefont{J.}~\bibnamefont{Gibbs}},
  \emph{\bibinfo{title}{Elementary principles in statistical mechanics}}
  (\bibinfo{publisher}{Yale University Press}, \bibinfo{address}{Yale},
  \bibinfo{year}{1902}), \bibinfo{note}{reprinted by Dover, New York, 1960}.

\bibitem[{\citenamefont{Huang}(1963)}]{Huang}
\bibinfo{author}{\bibfnamefont{K.}~\bibnamefont{Huang}},
  \emph{\bibinfo{title}{Statistical mechanics}} (\bibinfo{publisher}{John Wiley
  \& Sons}, \bibinfo{address}{Singapore}, \bibinfo{year}{1963}),
  \bibinfo{edition}{2nd} ed.

\bibitem[{\citenamefont{Campisi}(2005)}]{Campisi05}
\bibinfo{author}{\bibfnamefont{M.}~\bibnamefont{Campisi}},
  \bibinfo{journal}{Studies in History and Philosophy of Modern Physics}
  \textbf{\bibinfo{volume}{36}}, \bibinfo{pages}{275} (\bibinfo{year}{2005}).

\bibitem[{\citenamefont{Adib}(2004)}]{adib04}
\bibinfo{author}{\bibfnamefont{A.}~\bibnamefont{Adib}},
  \bibinfo{journal}{Journal of Statistical Physics}
  \textbf{\bibinfo{volume}{117}}, \bibinfo{pages}{581} (\bibinfo{year}{2004}).

\bibitem[{\citenamefont{Khinchin}(1949)}]{Khinchin}
\bibinfo{author}{\bibfnamefont{A.}~\bibnamefont{Khinchin}},
  \emph{\bibinfo{title}{Mathematical foundations of statistical mechanics}}
  (\bibinfo{publisher}{Dover}, \bibinfo{address}{New York},
  \bibinfo{year}{1949}).

\bibitem[{\citenamefont{Campa and Ruffo}(2006)}]{Campa06}
\bibinfo{author}{\bibfnamefont{A.}~\bibnamefont{Campa}} \bibnamefont{and}
  \bibinfo{author}{\bibfnamefont{S.}~\bibnamefont{Ruffo}},
  \bibinfo{journal}{Physica A} \textbf{\bibinfo{volume}{369}},
  \bibinfo{pages}{517} (\bibinfo{year}{2006}).

\bibitem[{\citenamefont{Gross and Kenney}(2005)}]{gross:224111}
\bibinfo{author}{\bibfnamefont{D.~H.~E.} \bibnamefont{Gross}} \bibnamefont{and}
  \bibinfo{author}{\bibfnamefont{J.~F.} \bibnamefont{Kenney}},
  \bibinfo{journal}{The Journal of Chemical Physics}
  \textbf{\bibinfo{volume}{122}}, \bibinfo{eid}{224111}
  (pages~\bibinfo{numpages}{8}) (\bibinfo{year}{2005}).

\bibitem[{\citenamefont{Thirring}(1970)}]{Thirring70}
\bibinfo{author}{\bibfnamefont{W.}~\bibnamefont{Thirring}},
  \bibinfo{journal}{Z. Physik} \textbf{\bibinfo{volume}{235}},
  \bibinfo{pages}{339} (\bibinfo{year}{1970}).

\bibitem[{\citenamefont{Naudts}(2005)}]{Naudts05}
\bibinfo{author}{\bibfnamefont{J.}~\bibnamefont{Naudts}},
  \bibinfo{journal}{Europhys. Lett.} \textbf{\bibinfo{volume}{69}},
  \bibinfo{pages}{719} (\bibinfo{year}{2005}).

\bibitem[{\citenamefont{Helmholtz}(1895)}]{Helmholtz}
\bibinfo{author}{\bibfnamefont{H.}~\bibnamefont{Helmholtz}},
  \emph{\bibinfo{title}{Principien der Statik monocyklischer Systeme, in
  ``Wissenshafltliche Abhandlungen'', vol III, p.142-162 and p.179-202; Studien
  zur Statik monocyklischer Systeme, in ``Wissenshafltliche Abhandlungen'', vol
  III, p.163-172 and p.173-178, Leipzig}} (\bibinfo{year}{1895}).

\bibitem[{\citenamefont{Cardin and Favretti}(2004)}]{Cardin04}
\bibinfo{author}{\bibfnamefont{F.}~\bibnamefont{Cardin}} \bibnamefont{and}
  \bibinfo{author}{\bibfnamefont{M.}~\bibnamefont{Favretti}},
  \bibinfo{journal}{Continuum Mechanics and Thermodynamics}
  \textbf{\bibinfo{volume}{16}}, \bibinfo{pages}{15} (\bibinfo{year}{2004}).

\bibitem[{\citenamefont{Uffink}(2006)}]{Uffink06}
\bibinfo{author}{\bibfnamefont{J.}~\bibnamefont{Uffink}},
  \emph{\bibinfo{title}{Compendium of the foundations of classical statistical
  physics}} (\bibinfo{year}{2006}),
  \urlprefix\url{http://philsci-archive.pitt.edu/archive/00002691}.

\bibitem[{\citenamefont{Hertz}(1910)}]{Hertz}
\bibinfo{author}{\bibfnamefont{P.}~\bibnamefont{Hertz}},
  \bibinfo{journal}{Annalen der Physik} \textbf{\bibinfo{volume}{33
  (Leipzig)}}, \bibinfo{pages}{225} (\bibinfo{year}{1910}).

\bibitem[{\citenamefont{M{\"u}nster}(1969)}]{Munster}
\bibinfo{author}{\bibfnamefont{A.}~\bibnamefont{M{\"u}nster}},
  \emph{\bibinfo{title}{Statistical thermodynamics}}, vol.~\bibinfo{volume}{1}
  (\bibinfo{publisher}{Springer Verlag}, \bibinfo{address}{Berlin},
  \bibinfo{year}{1969}).

\bibitem[{\citenamefont{Becker}(1969)}]{Becker69}
\bibinfo{author}{\bibfnamefont{R.}~\bibnamefont{Becker}},
  \emph{\bibinfo{title}{Theory of heat}}, vol.~\bibinfo{volume}{1}
  (\bibinfo{publisher}{Springer-Verlag}, \bibinfo{address}{Berlin},
  \bibinfo{year}{1969}), \bibinfo{edition}{2nd} ed.

\bibitem[{\citenamefont{Berdichevsky}(1997)}]{Berdichevsky}
\bibinfo{author}{\bibfnamefont{V.~L.} \bibnamefont{Berdichevsky}},
  \emph{\bibinfo{title}{Thermodynamics of chaos and order}}
  (\bibinfo{publisher}{Addison Wesley Longman}, \bibinfo{address}{Essex},
  \bibinfo{year}{1997}).

\bibitem[{\citenamefont{Uffink}(2001)}]{Uff01}
\bibinfo{author}{\bibfnamefont{J.}~\bibnamefont{Uffink}},
  \bibinfo{journal}{Studies in History and Philosophy of modern Physics}
  \textbf{\bibinfo{volume}{32}}, \bibinfo{pages}{305} (\bibinfo{year}{2001}).

\bibitem[{\citenamefont{Campisi}(2007)}]{Campisi07.4}
\bibinfo{author}{\bibfnamefont{M.}~\bibnamefont{Campisi}},
  \bibinfo{journal}{arXiv:0704.2567}  (\bibinfo{year}{2007}).

\bibitem[{\citenamefont{Tasaki}(2000)}]{Tasaki00}
\bibinfo{author}{\bibfnamefont{H.}~\bibnamefont{Tasaki}},
  \bibinfo{journal}{arXiv:cond-mat/0009206}  (\bibinfo{year}{2000}).

\bibitem[{\citenamefont{Pearson et~al.}(1985)\citenamefont{Pearson, Halicioglu,
  and Tiller}}]{Pearson85}
\bibinfo{author}{\bibfnamefont{E.~M.} \bibnamefont{Pearson}},
  \bibinfo{author}{\bibfnamefont{T.}~\bibnamefont{Halicioglu}},
  \bibnamefont{and} \bibinfo{author}{\bibfnamefont{W.~A.}
  \bibnamefont{Tiller}}, \bibinfo{journal}{Phys. Rev. A}
  \textbf{\bibinfo{volume}{32}}, \bibinfo{pages}{3030} (\bibinfo{year}{1985}).

\bibitem[{\citenamefont{Bianucci et~al.}(1995)\citenamefont{Bianucci, Mannella,
  West, and Grigolini}}]{BMWG}
\bibinfo{author}{\bibfnamefont{M.}~\bibnamefont{Bianucci}},
  \bibinfo{author}{\bibfnamefont{R.}~\bibnamefont{Mannella}},
  \bibinfo{author}{\bibfnamefont{B.}~\bibnamefont{West}}, \bibnamefont{and}
  \bibinfo{author}{\bibfnamefont{P.}~\bibnamefont{Grigolini}},
  \bibinfo{journal}{Physical Review E} \textbf{\bibinfo{volume}{51}},
  \bibinfo{pages}{3002} (\bibinfo{year}{1995}).

\bibitem[{\citenamefont{Landau and Lifshitz}(1960)}]{LandauI}
\bibinfo{author}{\bibfnamefont{L.}~\bibnamefont{Landau}} \bibnamefont{and}
  \bibinfo{author}{\bibfnamefont{E.}~\bibnamefont{Lifshitz}},
  \emph{\bibinfo{title}{Mechanics}} (\bibinfo{publisher}{Pergamon},
  \bibinfo{address}{Oxford}, \bibinfo{year}{1960}).

\bibitem[{\citenamefont{Galgani and Scott}(1972)}]{Galgani72}
\bibinfo{author}{\bibfnamefont{L.}~\bibnamefont{Galgani}} \bibnamefont{and}
  \bibinfo{author}{\bibfnamefont{A.}~\bibnamefont{Scott}},
  \bibinfo{journal}{Phys. Rev. Lett.} \textbf{\bibinfo{volume}{28}},
  \bibinfo{pages}{1173} (\bibinfo{year}{1972}).

\bibitem[{\citenamefont{Antoni and Ruffo}(1995)}]{Ruffo95}
\bibinfo{author}{\bibfnamefont{M.}~\bibnamefont{Antoni}} \bibnamefont{and}
  \bibinfo{author}{\bibfnamefont{S.}~\bibnamefont{Ruffo}},
  \bibinfo{journal}{Physical Review E} \textbf{\bibinfo{volume}{52}},
  \bibinfo{pages}{2361} (\bibinfo{year}{1995}).

\bibitem[{\citenamefont{Mukamel et~al.}(2005)\citenamefont{Mukamel, Ruffo, and
  Schreiber}}]{Mukamel05}
\bibinfo{author}{\bibfnamefont{D.}~\bibnamefont{Mukamel}},
  \bibinfo{author}{\bibfnamefont{S.}~\bibnamefont{Ruffo}}, \bibnamefont{and}
  \bibinfo{author}{\bibfnamefont{N.}~\bibnamefont{Schreiber}},
  \bibinfo{journal}{Phys. Rev. Lett.} \textbf{\bibinfo{volume}{95}},
  \bibinfo{pages}{240604} (\bibinfo{year}{2005}).

\end{thebibliography}

\end{document}